\documentclass[pra,twocolumn,superscriptaddress,10pt]{revtex4-1}
\usepackage{graphicx}
\usepackage{dcolumn}
\usepackage{color}
\usepackage{times}
\usepackage{bm}
\usepackage{amssymb}
\usepackage{amsmath}
\usepackage{dsfont}
\usepackage{float}
\usepackage{subfigure}
\usepackage{caption}
\usepackage{ragged2e}
\usepackage[colorlinks=true,citecolor=blue, linkcolor=blue, urlcolor=blue]{hyperref}
\begin{document}
\renewcommand{\thefootnote}{\fnsymbol {footnote}}

\title{Entropic uncertainty and quantum non-classicality of Unruh-Dewitt detectors in relativity}

\author{Yu-Kun Zhang}
\author{Li-Juan Li}
\author{Xue-Ke Song}
\author{Liu Ye}
\author{Dong Wang} \email{Corresponding author: dwang@ahu.edu.cn}
\affiliation{School of Physics \& Optoelectronic Engineering, Anhui University, Hefei 230601,  People's Republic of China}

\date{\today}

\begin{abstract}{
{An object moving with the acceleration will change  the temperature of environment around it,} because of the presence of the Unruh thermal effect. In this work, we investigate the impact of Unruh thermal noise on the quantum-memory-assisted {entropic} uncertainty   and quantum correlation regarding a pair of Unruh-Dewitt detectors. Specifically, we examine how the acceleration, the coupling strength between the external field and the detector, and the initial state affect the uncertainty and the system's quantum discord.  It turns out that the Unruh effect will result in the loss of the systemic quantumness and inflation of the uncertainty. Moreover, it is revealed that the uncertainty is  reversely correlated with the system's quantum discord. Thereby, it is believed that our investigations provide new insights into understanding the behavior of objects in the relativistic background.
}

\end{abstract}

\maketitle

\section{Introduction}
{It is widely recognized that} quantum mechanics and relativity theory are fundamental theories at explicit scales. The former belongs to the microscopic {domain} and the latter to the cosmological scale. 
Since the last century, scientists have worked to establish a link between
them. One of the important theories is quantum field theory, and some famous predictions are based on this theory. One of the most
interesting predictions is the Unruh effect \cite{PhysRevD.14.870,RevModPhys.80.787}, which describes a phenomenon that a uniformly
accelerated observer in flat space-time (also called Minkowski space-time) will {realize} that the temperature of vacuum space he stays
is rising. {However, as a static observer one cannot change the temperature.} 
So, the Unruh effect gives a new concept of the particle that is dependent on the observer \cite{PhysRevLett.91.180404,PhysRevLett.97.250502,PhysRevLett.95.120404, PhysRevLett.106.210502,PhysRevLett.110.113602,PhysRevLett.110.160501,PhysRevD.90.025032}, and gives another way to understand Hawking
radiation. Actually, both Unruh effect and Hawking radiation are caused by the vacuum fluctuations. 
{In the realm of quantum information}, under the background of the accelerated motions, {it is of basic importance to explore how the Unruh effect will affect quantum resources including  quantum correlation and uncertainty relations}.

One of the {most profound distinctions} between the classical physics and the quantum physics is regarded as the uncertainty relation, which was firstly proposed by Heisenberg 
\cite{Heisenberg1927}. For one single-particle system, one cannot obtain precise outcomes for two incompatible measurements $\hat{R}$ and $\hat{S}$, which can be expressed by the uncertainty relation
\begin{equation}
  \begin{aligned}
    \Delta \hat{R} \cdot \Delta \hat{S} \ge \frac{1}{2}|\langle\left[\hat{R},\hat{S}\right]\rangle|_\rho
    \label{1}
  \end{aligned}
\end{equation}
where  {$\Delta \hat{R} = \sqrt{\langle \hat{R}^{2}\rangle - \langle \hat{R}\rangle ^{2}}$
 and $\Delta \hat{S} = \sqrt{\langle \hat{S}^{2}\rangle - \langle \hat{S}\rangle ^{2}}$ are the  standard {deviations}, where $\langle \hat{R} \rangle = {\rm Tr} (\rho \hat{R})$ is the expectation value for the system's state $\rho$, and 
 $\left[\hat{R},\hat{S}\right] = \hat{R}\hat{S}-\hat{S}\hat{R}$ is the
commutator of observables $\hat{R}$ and $\hat{S}$. }
With the development of information theory, entropy is usually utilized to depict the uncertainty relation rather than the standard deviation.
Entropic uncertainty relation (EUR) was formulated by Deutsch \cite{PhysRevLett.50.631}, improved by Karus \cite{PhysRevD.35.3070} and proved by Maassen
and Uffink \cite{PhysRevLett.60.1103}, expressed as
\begin{equation}
  \begin{aligned}
    H(\hat{R}) + H(\hat{S}) \ge \log_{2}\frac{1}{c} = :q_{MU},
    \label{2}
  \end{aligned}
\end{equation}
where  $H\left({X}\right) = -\Sigma_{k} x_{k} \log_{2} x_{k}$ is the Shannon entropy of
the observable ${X} \in \left\{\hat{R},\hat{S}\right\} $, $x_{k}$ is the probability of the outcome $k$, and $q_{MU}$ denotes the incompatibility measure with $c = \max_{i,j} |\langle \hat{r}_{i}|\hat{s}_{j}\rangle|^{2}$, here $|\hat{r}_{i}\rangle$ and $|\hat{s}_{j}\rangle$ are the eigenstates of $R$ and $S$ respectively. For a composite system, 
Renes $et\ al$. \cite{PhysRevLett.103.020402} and Berta $et\ al$. \cite{Berta2010} {have} put forward quantum-memory-assisted
{entropic} uncertainty relations (QMA-EUR) regarding  a pair of arbitrary observables, which  can be mathematically written as the form of
\begin{align}
  S(\hat{Q}|{B})+S(\hat{R}|{B})\ge q_{MU}+S({A}|{B}),
  \label{3}
\end{align}
where $S(\hat{Q}|{B})=S(\hat{\rho}_{\hat{Q}B})-S(\hat{\rho}_{B})$ and $S(\hat{R}|{B})=S(\hat{\rho}_{\hat{R}B})-S(\hat{\rho}_{B})$ are von Neumann entropies of the post-measurement states. As a result, the quantum states can be given by
\begin{align}
  \begin{split}
    \hat\rho_{XB} & = \sum_{i}\left(|{x}_{i}\rangle \langle {x}_{i}|\otimes\mathbb{\hat{I}}\right) \hat{\rho}_{AB} \left(|{x}_{i}\rangle \langle {x}_{i}| \otimes \mathbb{\hat{I}}\right), \\
    \hat\rho_{ZB} & = \sum_{i}\left(|{z}_{i}\rangle \langle {z}_{i}|\otimes\mathbb{\hat{I}}\right) \hat{\rho}_{AB} \left(|{z}_{i}\rangle \langle {z}_{i}| \otimes \mathbb{\hat{I}}\right),
    \label{4}
  \end{split}
\end{align}
after performing two Pauli measurements $\hat{Z}$ and $\hat{X}$, where $\mathbb{\hat{I}}$ is the identity matrix, $|{x}_{i}\rangle$ and $|{z}_{i}\rangle$ are the eigenvectors of Pauli matrices. 
{As a matter of fact, there are much progress with respect to QMA-EUR} \cite{PhysRevA.75.022319,PhysRevA.86.042105,Pramanik2016,
PhysRevA.87.022314,PhysRevLett.110.020402,PhysRevA.89.022112,Zozor_2014,PhysRevA.91.042133,PhysRevA.93.062123,Huang2018,PhysRevA.102.012206,Ming2020,PhysRevA.106.062219,PhysRevE.106.054107,Li2022,PhysRevA.104.062204,Li2021,PhysRevA.101.032101,PhysRevE.109.064103}.
To understand QMA-EUR, we resort to {the} uncertainty  game between two legitimate players, say Alice and Bob. Both of them  in prior agree on two measurements $\hat{Q}$ and $\hat{R}$, and one of the players, say Bob,
prepares two particles  ${A}$ and ${B}$, in entangled state. Then, Bob sends particle  ${A}$  to another player Alice and keeps   $ {B}$   as the quantum memory. After that, Alice chooses either $\hat{Q}$ or $\hat{R}$ to measure, and records the outcome she obtains, meanwhile she shares her measurement's choice   with Bob via classical channel. The Bob's assignment   is to guess her result with the minimal uncertainty bounded by Eq. (\ref{3}). 

Compared with the classical counterpart, quantum resources can achieve the tasks that are difficult to depend on classical resources, such as, quantum coherence \cite{PhysRevLett.113.140401}, entanglement \cite{PhysRevLett.80.2245,Eisert_1999,PhysRevLett.103.160504}, quantum steering \cite{PhysRevA.87.062103} and Bell nonlocality \cite{PhysicsPhysiqueFizika.1.195,PhysRevA.75.022108,PhysRevLett.23.880} $et$ $al$. have been studied intensively in various quantum information processing tasks. For a state without entanglement,  there exists  non-zero quantum correlations. Particularly, Ollivier and Zurek \cite{PhysRevLett.88.017901} proposed the so-called quantum discord to quantify the quantum correlation, which can be mathematically expressed as
\begin{align}
    D\left(\hat{\rho}_{AB}\right)=I\left(\hat{\rho}_{AB}\right)-C\left(\hat{\rho}_{AB}\right)
    \label{new1}
\end{align}
where $I\left(\hat{\rho}_{AB}\right)$ represents {the} mutual information, written as 
\begin{align}
  I\left(\hat{\rho}_{AB}\right)=S\left(\hat{\rho}_{A}\right)+S\left(\hat{\rho}_{B}\right)-S\left(\hat{\rho}_{AB}\right),
  \label{5}
\end{align}
if we perform a POVM measurement of system $B$ with a set of measurement operators $\left\{\mathbb{\hat{M}}\right\}$, and we obtain a maximal quantity of mutual
information under one of the measuring bases, we call this maximal quantity classical correlation which can be written as  
\begin{align}
  C(\hat{\rho}_{AB})=\max_{\left \{\mathbb{\hat{M}}\right \}}\left\{S(\hat{\rho}_{A})-S(\hat{\rho}_{AB}|\mathbb{\hat{M}}_{B}^{i})\right\},
  \label{6}
\end{align}
the measured state $\hat{\rho}_{AB}^{i}$ has the form
\begin{equation}
  \begin{aligned}
    \hat{\rho}_{AB}^{i} = \frac{\mathrm{Tr}_{B}\left[\left(\mathbb{\hat{M}}_{A}\otimes \mathbb{\hat{M}}_{B}^{i}\right) \hat{\rho}_{AB} \left(\mathbb{\hat{I}}_{A}\otimes \mathbb{\hat{M}}_{B}^{i}\right)\right]}{P_{i}},
    \label{7}
  \end{aligned}
\end{equation}
where, $P_{i} = \mathrm{Tr} \left[\left(\mathbb{\hat{I}}_{A}\otimes \mathbb{\hat{M}}_{B}^{i}\right) \hat{\rho}_{AB} \left(\mathbb{\hat{I}}_{A}\otimes \mathbb{\hat{M}}_{B}^{i}\right)\right]$ represents the possibility
of the measuring outcome $i$. Substituting Eqs. (\ref{5}) and   (\ref{6}) into Eq. (\ref{new1}), the  quantum discord (QD) can be written as  
\begin{equation}
  \begin{aligned}
    D\left(\hat{\rho} _{AB}\right ) 
    = S\left (\hat{\rho}_{A}\right )-S\left (\hat{\rho} _{AB}\right )+\mathcal{M},
    \label{8}
  \end{aligned}
\end{equation}
where $\mathcal{M}=\min_{\left \{ \mathbb{M} \right \} }\left[S\left(\hat{\rho}_{AB}|\mathbb{\hat{M}}_{B}^{i}\right)\right ]$ with the conditional entropy $  S(\hat{\rho}_{AB}|\mathbb{\hat{M}}_{B}^{i})=\Sigma_{i} P_{i} S(\hat{\rho}_{AB}^{i})$ after the POVM measurement on
the particle $B$.

The outline of this article {is organized} as follows. In Sec. II, we briefly review the model  describing  two Unruh-Dewitt detectors that describes two Unruh-Dewitt detectors, along with the evolution of the entire system. In Sec. III, we observe the dynamics of the QMA-EUR and   QD with different parameters under Unruh thermal noises. At last, we end up our paper with  a concise conclusion.

\section{MODEL}
{Suppose there are two observers, Alice and Bob, each of whom owns an Unruh-Dewitt detector \cite{PhysRevD.29.1047} modeled by one two-level independent atom \cite{PhysRevD.68.085006,PhysRevA.80.032315,
  PhysRevA.81.062130,Tian_2012}, located in the Minkowski spacetime.}
 Alice owns the static detector and Bob owns the one that moves with uniform acceleration $a$ for one time duration $\Delta$. Alice
always switches her detector off, while Bob keeps his detector always be switched on. {Since  the world line functions of the system  can be written as follows}
\begin{equation}
  \begin{aligned}
    t\left(\tau\right) & = a^{-1} \sinh\left(a\tau\right), \\
    x\left(\tau\right) & = a^{-1} \cosh\left(a\tau\right), \\
    y\left(\tau\right) & = z\left(\tau\right) = 0,
    \label{10}
  \end{aligned}
\end{equation}
where $a$ is the acceleration of detector that Bob holds, and $\tau$ is the proper time of the moving detector. For simplification,  {$c=\hbar=\kappa_{B}=1$} are set hereafter.

We consider the system that bonds the detectors and external field together. 
We use {the states} $|\Psi_{A}\rangle$, $|\Psi_{B}\rangle$, $|0_{M}\rangle$ to represent the initial states of
Alice, Bob  and the Minkowski vacuum, respectively. The initial state of the detector-field system $|\hat{\Psi}_{0}^{ABM}\rangle$ can be written in the following form:
\begin{align}
  \begin{split}
    &|\hat{\Psi}_{0}^{ABM}\rangle  = |\hat{\Psi}_{AB}\rangle\otimes|\hat{0}_{M}\rangle,                                       \label{11}
  \end{split}
\end{align}
where $|\Psi_{AB}\rangle=\sin\theta|\hat{0}_{A}\rangle|\hat{1}_{B}\rangle + \cos\theta|\hat{1}_{A}\rangle|\hat{0}_{B}\rangle$ denotes the tensor product of Alice's and Bob's detector that in the initial state, and $|0_{M}\rangle$ signifies the initial state of the
external scalar field in the Minkowski vacuum. The total Hamiltonian of the system can be represented as
\begin{align}
  \hat{H}_{ABM}=\hat{H}_{A}+\hat{H}_{B}+\hat{H}_{M}+\hat{H}_{I}^{BM},
  \label{12}
\end{align}
where $\hat{H}_{A} = \Omega A^{\dagger} A$, $\hat{H}_{B} = \Omega B^{\dagger} B$ are the detectors' Hamiltonian and $\Omega$ denotes the energy gap between two energy levels of
the detectors. There is only one Bob's detector moving in the field, the detector will interact with the external {scalar field},
so the interaction Hamiltonian $\hat{H}_{I}^{BM}$ has
the following form
\begin{align}
  \hat{H}_{I}^{BM}\left(t\right)=\epsilon \left (t\right)\int_{\sum _{t}} \mathrm{d}^{3}\mathbf{x}
  \sqrt{-g } \phi\left (x\right )[\chi\left (\mathbf{x} \right ) \hat{B} +
  \bar{\chi}\left(\mathbf{x} \right )\hat{B}^{\dagger}  ],
  \label{13}
\end{align}
where $g \equiv \det\left(\mathfrak{g}_{ab}\right)$ with $\mathfrak{g}_{ab}$ being the Minkowski metric. Besides,   $\hat{B}(\hat{B}^{\dagger})$ is the annihilation
(creation) operators of Bob. Furthermore, $\chi\left(\mathbf{x}\right)=\left(\kappa \sqrt{2\pi}\right)^{-3}\exp\left(-\mathbf{x}^{2}/2 \kappa ^{2}\right)$
is a coupling function that vanishes outside a small volume around the detector. This is a Gaussian coupling function, and we are used to describe a dot detector
\cite{PhysRevD.68.085006} which only interacts with the neighbor scalar fields in Minkowski vacuum. {In the time duration of $\Delta$,} from $t_{0}$ to $t_{0}+\Delta$, 
the final state
\cite{PhysRevA.80.032315,PhysRevA.81.062130,Tian_2012,Wang2014,Wald:1995yp} of the detector-field system under the weak-coupling background \cite{PhysRevD.29.1047} can be written
as follows,
\begin{align}
 {|\hat{\Psi}_{t=t_{0}+\Delta}^{BM}\rangle=\left\{\hat{I}-i[\hat{\phi}\left(f\right)\hat{R}+\hat{\phi}^{\dagger}\left(f\right)\hat{R}^{\dagger}]\right\}|\hat{\Psi}_{t_{0}}^{BM}\rangle,} 
  \label{14}
\end{align}
and {$\hat{\phi}\left(f\right)$} is an operator about the distribution of the external scalar field, it can be written as,
\begin{equation}
  \begin{aligned}
    {\hat{\phi}\left(f\right)} & {= \int d^{4}x\sqrt{-g}\chi\left(x\right)f } \\
    & {= i\left[\hat{a}_{RI}\left(\overline{UE\bar{f}}\right)-\hat{a}_{RI}^{\dagger}\left(UEf\right)\right],} 
    \label{15}
  \end{aligned}
\end{equation}
{where, $f\equiv \epsilon\left(t\right)e^{-i \Omega t} \chi(\mathbf{x})$} is a compact support complex
function defined in the Minkowski space time, { $\hat{a}_{RI}\left(\bar{u}\right)$ and $\hat{a}_{RI}^{\dagger}\left(u\right)$ are the annihilation and creation operators of
$u = UEf$ modes \cite{PhysRevA.80.032315,PhysRevA.81.062130,Tian_2012,Wang2014,Wald:1995yp} respectively,}
{$U$ is an operator} taking the solutions of Klein-Gordon equation
in Rindler metric \cite{PhysRevA.80.032315,Wald:1995yp} with its positive-frequency part, $E$ is the difference between the advanced and retarded Green function,
{and ${\hat I}$ is} the identity operator.
{Then, the form of the final state can be represented as}
\begin{align}
    |\hat{\Psi}_{t}^{ABM}\rangle  =& |\hat{\Psi}_{t_{0}}^{ABM}\rangle+\sin\theta|\hat{0}_{A}\rangle|\hat{0}_{B}\rangle \otimes \left[\hat{a}_{RI}^{\dagger}\left(\lambda\right)|\hat{0}_{M}\rangle\right] \nonumber \\
& + \cos\theta|\hat{1}_{A}\rangle|\hat{1}_{B}\rangle \otimes \left[\hat{a}_{RI}\left(\bar{\lambda}\right)|\hat{0}_{M}\rangle\right],
\label{16}
\end{align}
in terms of Eqs. (\ref{8}) and  (\ref{12}).  After tracing out the degree of freedom of the external field
{$\hat{\phi}\left(f\right)$},  the density matrix reflecting the state of the detectors can be expressed  as 
\begin{equation}
  \hat{\rho}_{AB} =
  \begin{pmatrix}
    \gamma & 0                                    & 0                                    & 0     \\
    0      & 2\alpha \sin^{2} \left(\theta\right) & \alpha \sin \left(2\theta \right )   & 0     \\
    0      & \alpha \sin \left(2\theta \right )   & 2\alpha \cos^{2} \left(\theta\right) & 0     \\
    0      & 0                                    & 0                                    & \beta
    \label{17}
  \end{pmatrix}.
\end{equation}
Within the above, $\alpha,\beta$ and $\gamma$ are given by
\begin{equation}
  \begin{aligned}
    \alpha & = \frac{1-q}{2 \left(1-q\right) + 2 \nu^{2} \left(\sin^{2}\theta + q \cos^{2} \theta\right)},                      \\
    \beta  & = \frac{\nu^{2} q \cos^{2}\theta}{2 \left(1-q\right) + 2 \nu^{2} \left(\sin^{2}\theta + q \cos^{2} \theta\right)}, \\
    \gamma & = \frac{\nu^{2} \sin^{2}\theta}{2 \left(1-q\right) + 2 \nu^{2} \left(\sin^{2}\theta + q \cos^{2} \theta\right)},
    \label{18}
  \end{aligned}
\end{equation}
respectively. In these parameters, $q= {\rm exp}{(-\frac{2\pi\Omega}{a})}$ is   related to the acceleration $a$ of the detector, and we have $q\to1$ in the limit  of $a\to\infty$. Additionally, the effective coupling strength $\nu$ satifies $\nu^{2} \equiv \left\| \lambda \right\|^{2} = \frac{\epsilon^{2}\Omega\Delta}{2\pi}
  e^{-\Omega^{2}\kappa^{2}}$ \cite{PhysRevA.80.032315,Tian_2012,Wald:1995yp}, {with $\nu^{2}\ll 1$ and $\Omega^{-1}\ll\Delta$.}

\section{DYNAMICS Behaviors OF  QD AND QMA-EUR FOR THE DETECTOR MODEL}
{In order to reveal the quantumness of the two Unruh-Dewitt detectors,} we here can resort to 
two Pauli {measurements}  {$\hat{X}$ and $\hat{Z}$}. As a result,  the specific forms of {$\hat{\rho}_{\hat{X}B}$} and 
{$\hat{\rho}_{\hat{Z}B}$} can be offered as 
\begin{equation}
  \begin{aligned}
    {\hat{\rho}_{\hat{X}B}} =
    \begin{pmatrix}
      \Xi     & 0       & 0       & \Lambda \\
      0       & \Theta  & \Lambda & 0       \\
      0       & \Lambda & \Xi     & 0       \\
      \Lambda & 0       & 0       & \Theta
    \end{pmatrix}, \ \
    {\hat{\rho}_{\hat{Z}B}} =
    \begin{pmatrix}
      \gamma & 0    & 0    & 0     \\
      0      & \Phi & 0    & 0     \\
      0      & 0    & \Psi & 0     \\
      0      & 0    & 0    & \beta
    \end{pmatrix},
  \end{aligned}
  \label{19}
\end{equation}
according to Eq. (\ref{4}). Where, $\Xi=\frac{\gamma}{2}+\alpha\cos^{2}\theta$, $\Theta=\frac{\beta}{2}+\alpha\sin^2\theta$, $\Lambda=\alpha\cos\theta\sin\theta$,
$\Phi=2\alpha\sin^{2}\theta$, and $\Psi=2\alpha\cos^{2}\theta$ respectively. Furthermore, the reduced density matrix of Bob's detector by tracing out the degree of freedom of Alice's detector is written as
\begin{equation}
  \hat{\rho}_{B} =
  \begin{pmatrix}
    \gamma+2\alpha\cos^{2}\theta & 0                           \\
    0                            & \beta+2\alpha\sin^{2}\theta \\
  \end{pmatrix}.
  \label{20}
\end{equation}
Upon these matrices, the explicit form of the entropic uncertainty, i.e, the left-hand side (LHS) of Eq. (\ref{3}), can be given by
\begin{equation}
  \begin{aligned}
   {S\left(\hat{X}|B\right)+S\left(\hat{Z}|B\right)}   
   =  & -\Sigma_{i}\lambda_{i}\log_{2}\left(\lambda_{i}\right) \\
                                            & -\Sigma_{j}\epsilon_{j}\log_{2}\left(\epsilon_{i}\right)  \\
                                            & +2\Sigma_{k}\mu_{k}\log_{2}\left(\mu_{k}\right),
  \end{aligned}
  \label{21}
\end{equation}
where $\left\{\lambda_{i}|i=1,2,3,4\right\}$ and $\left\{\epsilon_{j}|j=1,2,3,4\right\}$ are eigenvalues of the matrices {$\hat{\rho}_{{\hat{X}}B}$} and 
{$\hat{\rho}_{{\hat{Z}}B}$} respectively, and $\left\{
  \mu_{k}|k=1,2\right\}$ is the eigenvalue of the matrix {$\hat{\rho}_{B}$}.

\begin{figure}[h]
  \subfigure{\includegraphics[width=0.45\textwidth]{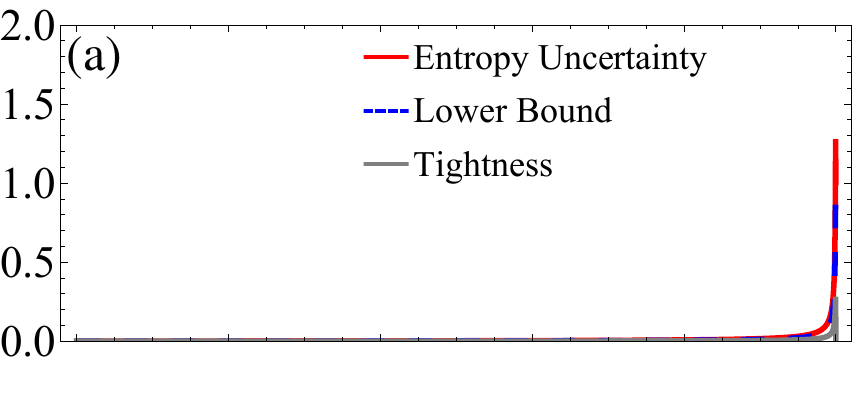}}\label{fig:subfig1}
  \vskip -0.62cm
  \subfigure{\includegraphics[width=0.45\textwidth]{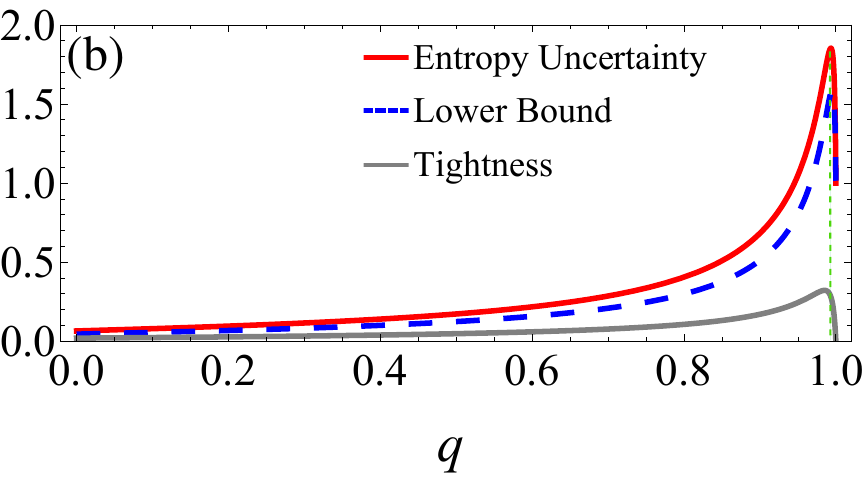}\label{fig:subfig2}}
  \caption{\justifying The dynamics of the uncertainty, its bound and tightness with the parameterized acceleration $q$ at different coupling strength $\nu$. Graph (a):  $\nu$ = 0.01 and Graph (b): $\nu$ = 0.1. And  the {state's} parameter  $\theta$ = $\pi/4$. In the plots, the red solid line represents the entropy uncertainty, the blue dashed line represents its lower bound, and the grey solid line represents the tightness of the uncertainty bound.}
  \label{fig1}
\end{figure}

 Fig. \ref{fig1}   illustrates the dynamic evolution of the uncertainty, its bound and its tightness (i.e., the difference of uncertainty and its bound) of QMA-EUR as a function of the parameterized acceleration $q$. It is evident that  the measured uncertainty for the two detectors generally increases, as the parameterized acceleration $q$ increases. Interestingly, as $q\to 1$ (i.e., $a\to \infty$), the uncertainty exhibits a sharp decline for extremely large acceleration, { and declines to 1 at $q=1$,} as shown in Fig. \ref{fig1}(a). It is observed that, at relatively low coupling strengths $\nu$ of between the field and detector, the magnitude of the uncertainty is immune to the parameterized  acceleration $q$, and  
 becomes appreciable only when $a\to \infty$; As to the strong coupling strength $\nu$, the effect of acceleration emerges more proactively, as displayed in Fig. \ref{fig1}(b). Additionally, the tightness of the uncertainty bound is plotted as well, and is equal or more than $0$ always, {which readily supports that Eq. (\ref{3}) is held}. Moreover, the weaker coupling strength between the detectors and the external field will result in the higher tightness. 

Next, we turn to study the system's quantum discord defined as Eq. (\ref{new1}), in order to examine the quantumness of the system of interest.   Resorting to Ref. \cite{Wang_2011},   we perform a POVM measurement on subsystem $B$ using a set of measuring operators $\left\{\mathbb{\hat{M}}_{k}=|\hat{M}_{k}\rangle \langle \hat{M}_{k}|,k=1,2\right\}$ to minimize the conditional entropy
of subsystem $A$, where
$|\hat{M}_{1}\rangle \equiv \cos\left(\frac{\eta}{2}\right) |\hat{1}\rangle + e^{i\zeta} \sin\left(\frac{\eta}{2}\right) |\hat{0}\rangle$ and
$|\hat{M}_{2}\rangle \equiv \sin\left(\frac{\eta}{2}\right) |\hat{1}\rangle - e^{i\zeta} \cos\left(\frac{\eta}{2}\right) |\hat{0}\rangle$ with
$0 \le \eta \le \frac{\pi}{2}$ and $0 \le \zeta \le 2\pi$.
The probability $p_{k}$   related to the result $k$ and the eigenvalues of the corresponding $\hat{\rho}_{k}$   are given by
\begin{equation}
  \begin{aligned}
     & p_{k}                               = \frac{1}{2}\left[1+(-1)^{k}\cos\theta\left(1-2\rho_{11}-2\rho_{33}\right)\right], \\
     & \lambda_{\pm}\left({\hat{\rho}_{k}}\right)  = \frac{1}{2}\left(1 \pm \frac{1}{p_{k}}\sqrt{\xi_{k}}\right),
  \end{aligned}
  \label{22}
\end{equation}
respectively, where $\xi_{k}$ is euqal to 
\begin{equation}
  \begin{aligned}
    \xi_{k}  = & \frac14 {\left[ 1 - 2\left(\rho_{33} + \rho_{44}\right) + \left(-1\right)^{k} \cos \left(1 - 2\rho_{11} - \rho_{44}\right) \right]^{2}}    \\
            & + \sin^{2}\eta \left[\left(\rho_{14}\right)^{2} + \left(\rho_{23}\right)^{2} - 2|\rho_{14}\rho_{23}| \sin\left(2\zeta+\phi\right) \right],
  \end{aligned}
  \label{23}
\end{equation}
where $\phi$ satisfies $\cos\phi = \frac{\mathrm{Im}\left(\rho_{14}\rho_{32}\right)}{|\rho_{14}\rho_{23}|}$ and $\sin\phi = \frac{\mathrm{Re}\left(\rho_{14}\rho_{32}\right)}{|\rho_{14}
    \rho_{23}|}$, with $\left\{\rho_{ij}|i=1,2,3,4;j=1,2,3,4\right\}$ is the element of density matrix $\hat{\rho}_{AB}$. The entropy of $\hat{\rho}_{k}$ can be written as $S\left(\hat{\rho}_{k}\right) = H\left(\lambda_{+}\left(\hat{\rho}_{k}\right)\right)$ where the binary entropy $H\left(\Lambda\right) \equiv -\Lambda\log_{2}
  \Lambda - \left(1-\Lambda\right) \log_{2} \left(1-\Lambda\right)$. As a result,  the conditional entropy can be expressed as
\begin{equation}
  \begin{aligned}
    S\left({\hat{\rho}_{A|B}}\right) = p_{1}S\left({\hat{\rho}_{1}}\right) + p_{2}S\left({\hat{\rho}_{2}}\right),
    \label{24}
  \end{aligned}
\end{equation}
when
\begin{equation}
  \begin{aligned}
    \Lambda = \frac{1 + \sqrt{\left[1 - 2\left(\rho_{33}+\rho_{44}\right)\right]^{2} + 4\left(|\rho_{14}|+|\rho_{23}|\right)^{2}} }{2}.
  \end{aligned}
  \label{25}
\end{equation}
In order to compute the minimum of the conditional entropy  $S\left(\hat{\rho}_{A|B}\right)$, Eq. (\ref{24}) can be
derived  with respect to $\eta$ and $\zeta$ as follows:
\begin{equation}
  \begin{aligned}
    \frac{\partial S\left(\hat{\rho}_{A|B}\right)}{\partial \eta}  & = 0,  \\
    \frac{\partial S\left(\hat{\rho}_{A|B}\right)}{\partial \zeta} & = 0.
    \label{26}
  \end{aligned}
\end{equation}
Then,  we can obtain two minimal values $\Gamma_{1}
  = H\left(\Lambda\right)$ and $\Gamma_{2} = -\Sigma_{i}\rho_{ii}\log_{2}\rho_{ii}-H\left(\rho_
  {11}+\rho_{33}\right)$ depending on diagonal elements of $\hat{\rho}_{AB}$. Thus, the minimum of the conditional entropy can be expressed as
\begin{equation}
  \begin{aligned}
    \min_{\left\{\mathbb{\hat{M}}\right\}}S\left(\hat{\rho}_{AB}|\mathbb{\hat{M}}\right) = \min\left(\Gamma_{1},\Gamma_{2}\right).
  \end{aligned}
  \label{27}
\end{equation}
Combining Eqs. (\ref{6}) and (\ref{8}), the explicit expressions of classical correlation and quantum discord can be worked out accordingly.
\begin{figure}[h]
  \subfigure
  {\includegraphics[width=0.225\textwidth]{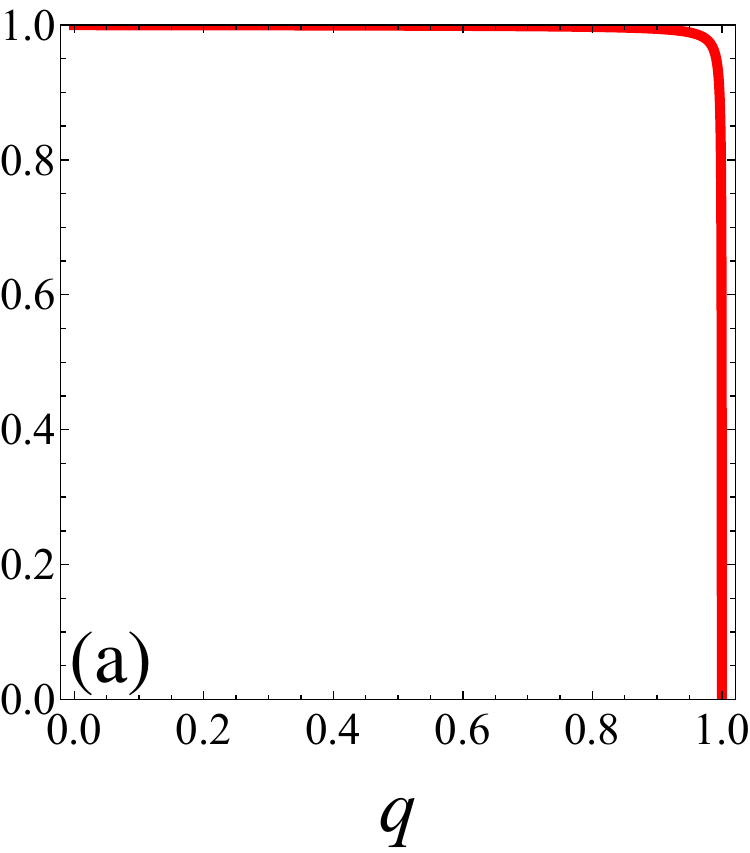}\label{fig:subfig3}}
  \subfigure\ \ 
  {\includegraphics[width=0.225\textwidth]{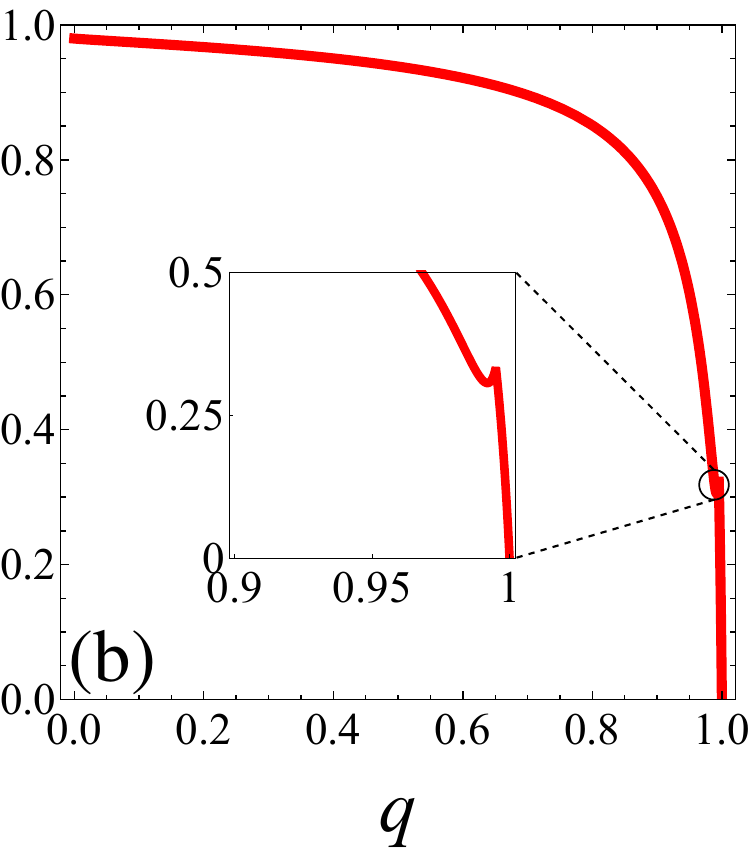}\label{fig:subfig4}}
  \caption{\justifying The dynamics of quantum discord ($D\left(\hat{\rho} _{AB}\right )$) versus the parameterized acceleration $q$ for the different coupling strength $\nu$. And the {state's} parameter $\theta=\pi/4$ is set. Graph (a): $\nu$ = 0.01 and Graph (b): $\nu$ = 0.1.}
  \label{fig2}
\end{figure}

Fig. \ref{fig2} has drawn the quantum discord (i.e., $D\left(\hat{\rho} _{AB}\right )$) as a function of the parameterized acceleration $q$ with the different couplings $\nu=0.01, 0.1$. It is interesting to find: (i) there exists an inverse correlation between {entropic} uncertainty and quantum discord, that is, an escalation in {the entropic} uncertainty tends to be accompanied by the decreasing {the} quantum discord, by compared with Figs. \ref{fig1} and \ref{fig2}. (ii) The acceleration will reduce the quantum discord. In other words, the larger acceleration will degrade the   quantum correlation of the detectors' system, due to destroying the system's purity. (iii) The coupling strength of between the field and detector has negative effect in the non-classicality  of the system. Specifically, the system's quantumness is insensitive to the  acceleration with respect to weak coupling regions, as shown Fig. \ref{fig1}(a). Besides, there is a singular point for quantum discord in Fig. \ref{fig1}(b). This is because of taking the minimum of the conditional entropy {as expressed in} Eq. (\ref{27}). Incidentally, whatever  $\nu$ is, quantum discord always declines to $0$ for $a\to \infty$. This occurrence underscores the pronounced impact of Unruh thermal noise on quantum systems subjected to maximum acceleration, leading to a precipitous degradation of quantum resources.

\begin{figure}[t]
  \subfigure
  {\includegraphics[width=0.23\textwidth]{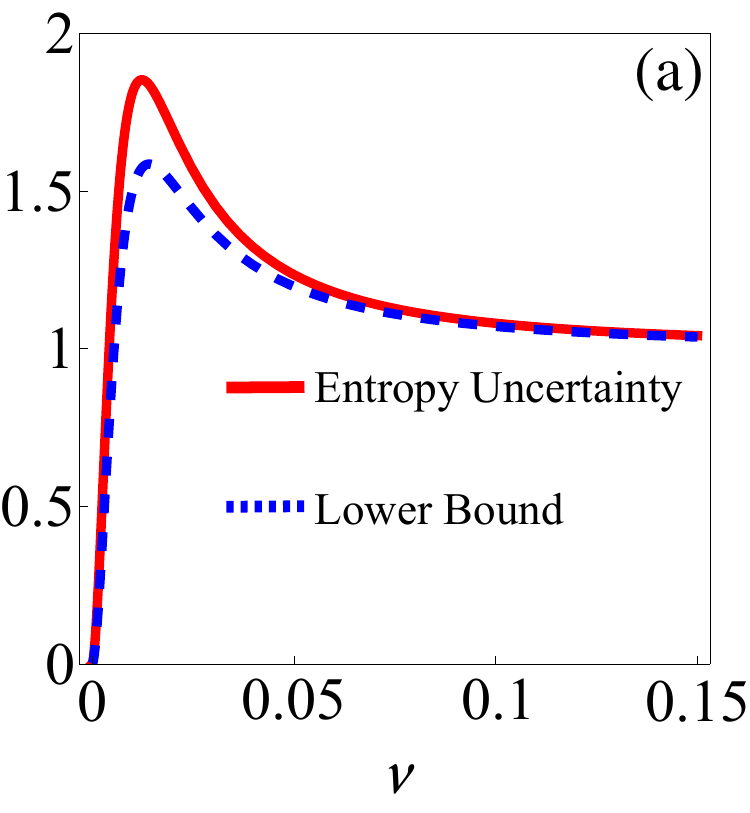}\label{fig:subfig5}}
  \subfigure\ \
  {\includegraphics[width=0.23\textwidth]{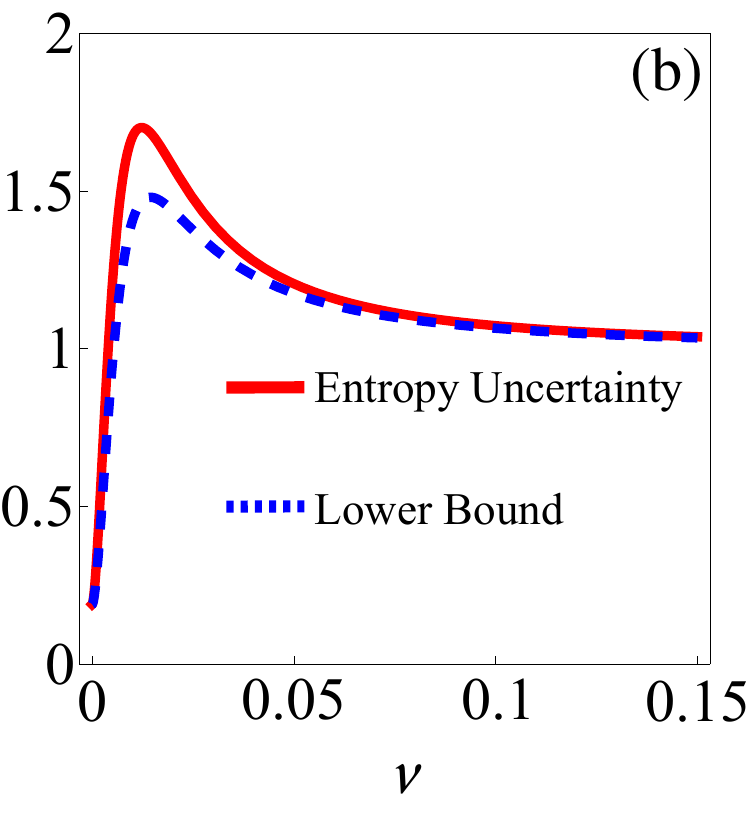}\label{fig:subfig6}}
  \caption{\justifying The dynamics of uncertainty with the coupling strength $\nu$. Graph (a): $\theta=\pi/4$ and Graph (b): $\theta=\pi/6$. And the parameterized acceleration $q=0.999$. In both plots, the red solid line represents the entropy uncertainty, the blue dashed line represents its lower bound.}
  \label{fig3}
\end{figure}

Additionally, Fig. \ref{fig3} demonstrates   how  the coupling strength $\nu$ between detector and field {impacts the dynamic evolution} of the entropy uncertainty when the high acceleration scenario is considered $(q = 0.999)$. As illustrated in the figure, 
the uncertainty and its lower bound will firstly increase and then gradually decrease as the growth of the coupling strength.  
 Notably, it should be emphasized that the
diverse initial states can affect the magnitude of the uncertainty  significantly, manifesting both the external field  and the initial state are pivotal factors {in determining} the dynamics of the {entropic} uncertainty under relativity.

\begin{figure}[tbh]
  \subfigure
  {\includegraphics[width=0.23\textwidth]{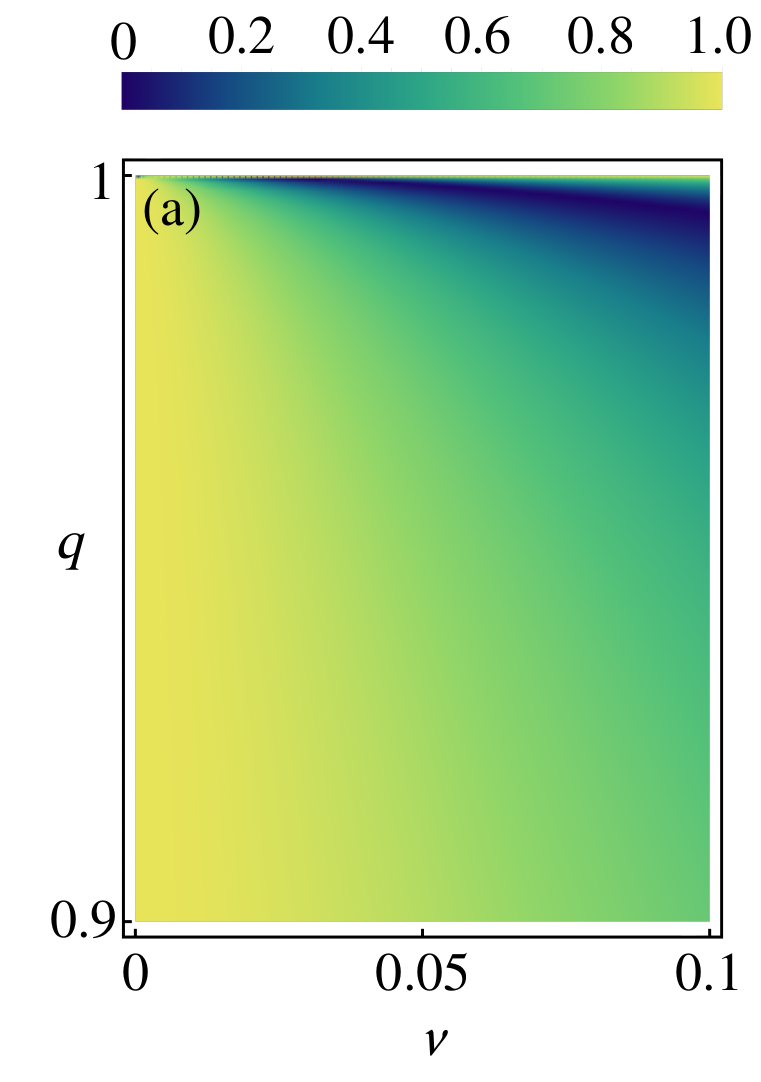}\label{fig:subfig7}}\ \
  \subfigure
  {\includegraphics[width=0.23\textwidth]{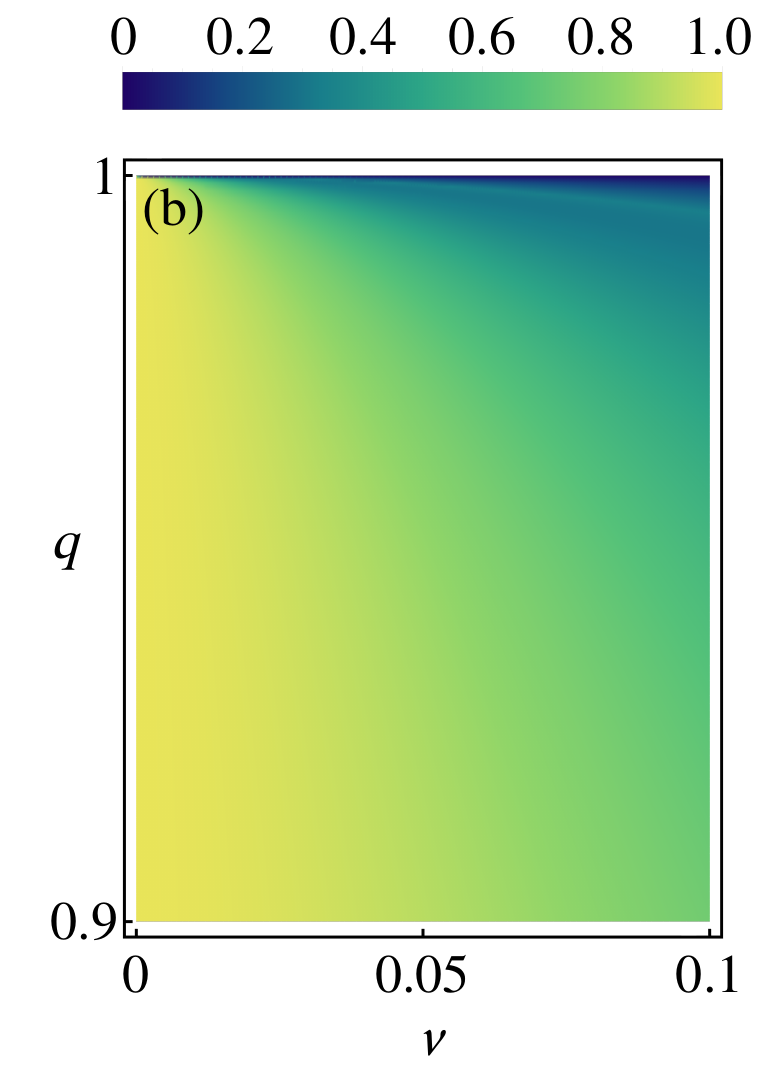}\label{fig:subfig8}}
  \caption{\justifying Contour of classical correlation {(Graph (a))} and quantum discord {(Graph (b))} with parameterized acceleration $q$  and the coupling strength $\nu$, and the {state's} parameter is set at $\theta=\pi/4$. }
  \label{fig4}
\end{figure}

In order to better understand  the information distribution including quantum and classical parts,
Fig. \ref{fig4} illustrates the dynamical evolution of classical correlation and quantum discord between the two detectors, by considering the composite influence of the parameterized acceleration and coupling strength. Following the figure, both the quantum discord and classical correlation will decrease as the coupling strength $\nu$; intriguingly, the classical correlation  will have a non-monotonic variation as the growing $q$, i.e., decreasing up to $0$ and then increasing, while the quantum discord will degrade as the growing  $q$. This behavior may attribute to the information redistribution in the case of {relatively} large acceleration, namely, there might exhibit the tradeoff between quantum and classical information in the current scenario. 

\begin{figure}[h]
  \centering
  \subfigure
  {\includegraphics[width=0.23\textwidth]{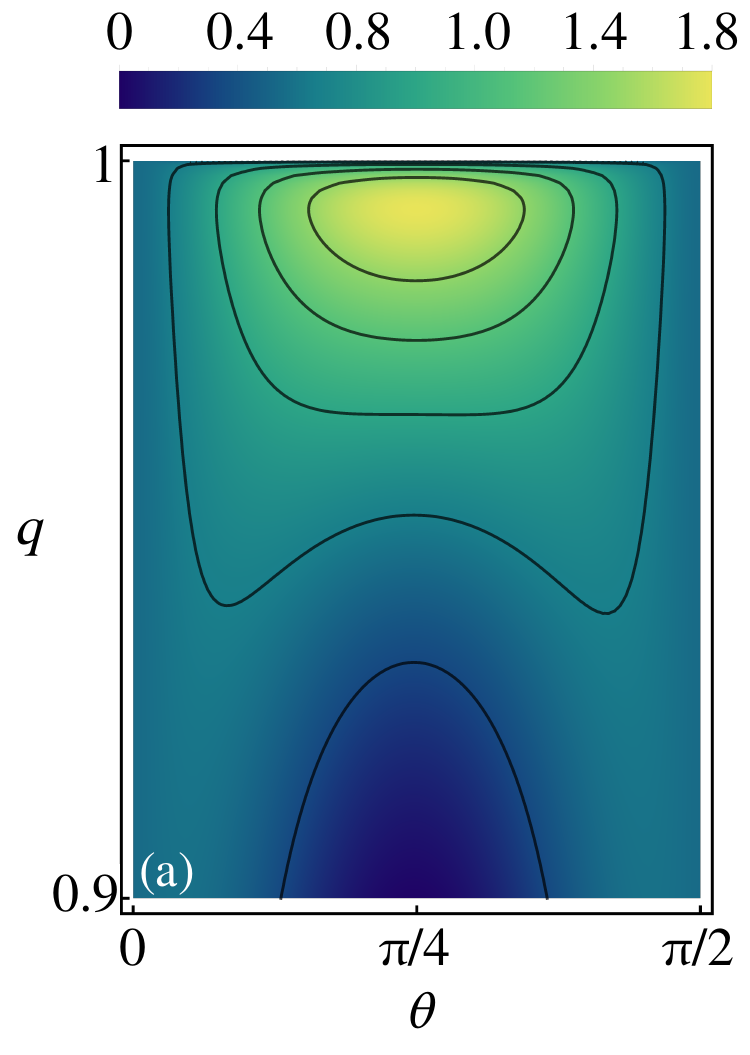}\label{fig:subfig9}}\ \
  \subfigure
  {\includegraphics[width=0.23\textwidth]{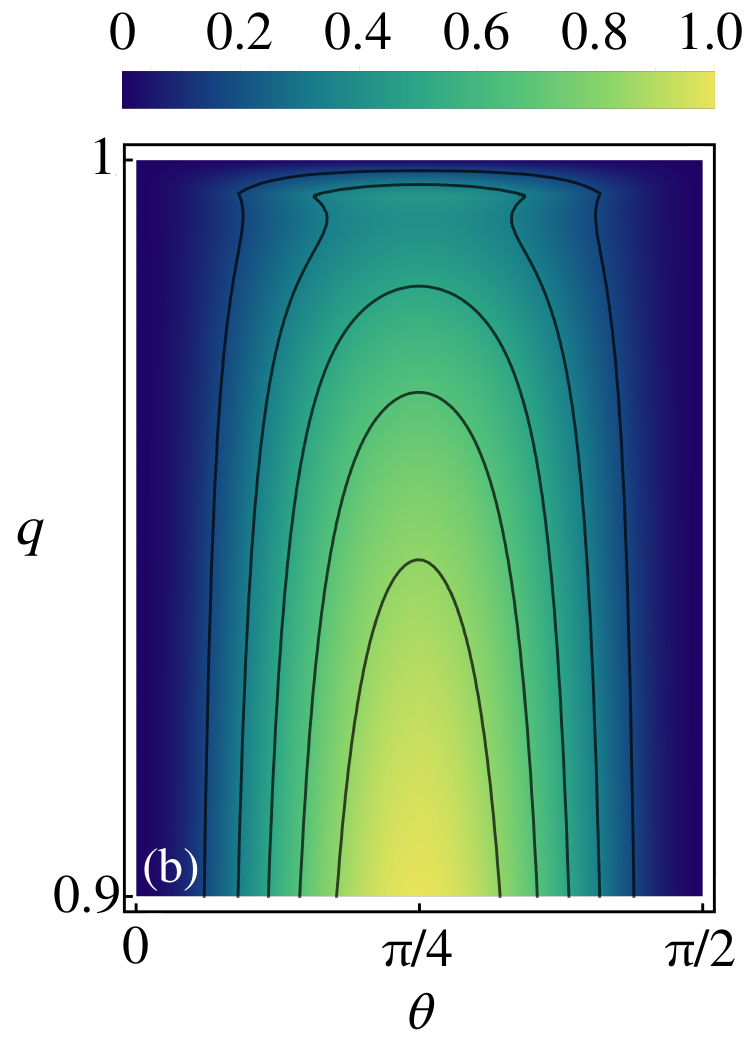}\label{fig:subfig10}}
  \caption{\justifying Contour of the entropy uncertainty  {(Graph (a))}  and the quantum discord {(Graph (b))}  with parameterized acceleration $q$ and the {state's} parameter $\theta$, with the coupling strength is set to $\nu=0.1$.}
  \label{fig5}
\end{figure}

To further elucidate the relationship between the uncertainty and quantum discord, we plot both of them as functions of the acceleration $q$ and initial state's parameter $\theta$ in  Fig \ref{fig5}. It is straightforward to obtain that: (i) the uncertainty and quantum discord are both symmetric with respect to the state's parameter $\theta=\pi/4$; (ii) {the uncertainty is anti-correlated with the quantum correlation, which essentially is in agreement with the result obtained before.}


\section{CONCLUSIONS}
In summary, we have investigated the {behaviors}  of quantum-memory-assisted entropy uncertainty and quantum non-classicality between two entangled detectors in the context of Minkowski space-time. It has been revealed that the Unruh effect, the coupling strength between the external field and the detector, and the initial state's parameter will influence the measured uncertainty and quantumness of the system of interest. Several interesting results are obtained as: (i) The growing acceleration will yield initial inflation and subsequent reduction of the uncertainty, and generally can reduce the system's quantum discord; (ii) The   coupling strength between the external field and the detector will destroy the system's quantum discord, and the uncertainty will firstly increase and then decrease into a stable value   as growth of the coupling strength; (iii) The uncertainty and quantum discord are symmetric with regard to the state's parameter $\theta=\pi/4$; (iv) {The uncertainty is nearly anti-correlated with the quantum correlation.}
 With these in mind, it is claimed that our investigations shed light on the entropy-based uncertainty relation and quantum non-classicality {of the Unruh-Dewitt detectors,} and is of fundamental importance to perspective quantum information processing in the framework
  of relativity.

\begin{acknowledgements}
This work was supported by the National Science Foundation of China (Grant nos. 12475009, 12075001,  and 62471001), Anhui Provincial Key Research and Development Plan (Grant No. 2022b13020004), Anhui Province Science and Technology Innovation Project (Grant No. 202423r06050004) and Anhui Provincial University Scientific Research Major Project (Grant No. 2024AH040008) .
\end{acknowledgements}
\vfil


\begin{references}

\bibitem {PhysRevD.14.870}
W. G. Unruh, \href{https://link.aps.org/doi/10.1103/PhysRevD.14.870}
{Phys. Rev. D. {\bf 14}, 870 (1976)}.

\bibitem {RevModPhys.80.787}
L. C. B. Crispino, A. Higuchi, and G. E. A. Matsas, 
\href{https://link.aps.org/doi/10.1103/RevModPhys.80.787}
{Rev. Mod. Phys. {\bf 80}, 787 (2008)}.

\bibitem {PhysRevLett.91.180404} 
P. M. Alsing and G. J. Milburn, \href{https://link.aps.org/doi/10.1103/PhysRevLett.91.180404}
{Phys. Rev. Lett. {\bf 91}, 180404 (2003)}.

\bibitem {PhysRevLett.97.250502} 
L. Lamata, M. A. Martin-Delgado, and E. Solano, \href{https://link.aps.org/doi/10.1103/PhysRevLett.97.250502}
{Phys. Rev. Lett. {\bf 97}, 250502 (2006)}.

\bibitem {PhysRevLett.95.120404} 
I. Fuentes-Schuller and R. B. Mann, \href{https://link.aps.org/doi/10.1103/PhysRevLett.95.120404}
{Phys. Rev. Lett. {\bf 95}, 120404 (2005)}.

\bibitem {PhysRevLett.106.210502} 
T. G. Downes, I. Fuentes, and T. C. Ralph, \href{https://link.aps.org/doi/10.1103/PhysRevLett.106.210502}
{Phys. Rev. Lett. {\bf 106}, 210502 (2011)}.

\bibitem {PhysRevLett.110.113602} 
N. Friis, A. R. Lee, K. Truong, C. Sab\'{\i}n, E. Solano, G. Johans-
son, and I. Fuentes,
\href{https://link.aps.org/doi/10.1103/PhysRevLett.110.113602}
{Phys. Rev. Lett. {\bf 110}, 113602 (2013)}.

\bibitem {PhysRevLett.110.160501} 
E. Mart\'{\i}n-Mart\'{\i}nez, D. Aasen, and A. Kempf, \href{https://link.aps.org/doi/10.1103/PhysRevLett.110.160501}
{Phys. Rev. Lett. {\bf 110}, 160501 (2013)}.

\bibitem {PhysRevD.90.025032} 
J. Wang, J. Jing, and H. Fan, \href{https://link.aps.org/doi/10.1103/PhysRevD.90.025032}
{Phys. Rev. D {\bf 90}, 025032 (2014)}.

\bibitem {Heisenberg1927} 
W. Heisenberg,
\href{https://doi.org/10.1007/BF01397280}
{Z. Physik {\bf 43}, 172 (1927)}.

\bibitem {PhysRevLett.50.631} 
D. Deutsch, \href{https://link.aps.org/doi/10.1103/PhysRevLett.50.631}
{Phys. Rev. Lett. {\bf 50}, 631 (1983)}.

\bibitem {PhysRevD.35.3070} 
K. Kraus, 
\href{https://link.aps.org/doi/10.1103/PhysRevD.35.3070}
{Phys. Rev. D {\bf 35}, 3070 (1987)}.

\bibitem {PhysRevLett.60.1103} 
H. Maassen and J. B. M. Uffink, 
\href{https://link.aps.org/doi/10.1103/PhysRevLett.60.1103}
{Phys. Rev. Lett. {\bf 60}, 1103 (1988)}.

\bibitem {PhysRevLett.103.020402} 
J. M. Renes and J.-C. Boileau, \href{https://link.aps.org/doi/10.1103/PhysRevLett.103.020402}
{Phys. Rev. Lett. {\bf 103}, 020402 (2009)}.

\bibitem {Berta2010} 
M. Berta, M. Christandl, R. Colbeck, J. M. Renes, and R. Ren-
ner,
\href{https://doi.org/10.1038/nphys1734}
{Nat. Phys. {\bf 6}, 659 (2010)}.

\bibitem {PhysRevA.75.022319} 
M. A. Ballester and S. Wehner, \href{https://link.aps.org/doi/10.1103/PhysRevA.75.022319}
{Phys. Rev. A {\bf 75}, 022319 (2007)}.

\bibitem {PhysRevA.86.042105} 
A. K. Pati, M. M. Wilde, A. R. U. Devi, A. K. Rajagopal, and
Sudha,
\href{https://link.aps.org/doi/10.1103/PhysRevA.86.042105}
{Phys. Rev. A {\bf 86}, 042105 (2012)}.

\bibitem {Pramanik2016} 
 T. Pramanik, S. Mal, and A. S. Majumdar, 
\href{https://doi.org/10.1007/s11128-015-1187-6}
{Quantum Inf. Process. {\bf 15}, 981 (2016)}.

\bibitem {PhysRevA.87.022314} 
 M.-L. Hu and H. Fan, \href{https://link.aps.org/doi/10.1103/PhysRevA.87.022314}
{Phys. Rev. A {\bf 87}, 022314 (2013)}.

\bibitem {PhysRevLett.110.020402} 
T. Pramanik, P. Chowdhury, and A. S. Majumdar, \href{https://link.aps.org/doi/10.1103/PhysRevLett.110.020402}
{Phys. Rev. Lett. {\bf 110}, 020402 (2013)}.

\bibitem {PhysRevA.89.022112} 
P. J. Coles and M. Piani, \href{https://link.aps.org/doi/10.1103/PhysRevA.89.022112}
{Phys. Rev. A {\bf 89}, 022112 (2014)}.

\bibitem {Zozor_2014} 
S. Zozor, G. M. Bosyk, and M. Portesi, \href{https://dx.doi.org/10.1088/1751-8113/47/49/495302}
{J. Phys. A: Math. Theor. {\bf 47}, 495302 (2014)}.

\bibitem {PhysRevA.91.042133} 
S. Liu, L.-Z. Mu, and H. Fan,  \href{https://link.aps.org/doi/10.1103/PhysRevA.91.042133}
{Phys. Rev. A {\bf 91}, 042133 (2015)}.

\bibitem {PhysRevA.93.062123} 
F. Adabi, S. Salimi, and S. Haseli, \href{https://link.aps.org/doi/10.1103/PhysRevA.93.062123}
{Phys. Rev. A {\bf 93}, 062123 (2016)}.

\bibitem {Huang2018} 
J.-L. Huang, W.-C. Gan, Y. Xiao, F.-W. Shu, and M.-H. Yung,
\href{https://doi.org/10.1140/epjc/s10052-018-6026-3}
{Eur. Phys. J. C {\bf 78}, 545 (2018)}.

\bibitem {PhysRevA.102.012206} 
F. Ming, D. Wang, X.-G. Fan, W.-N. Shi, L. Ye, and J.-L. Chen,
\href{https://link.aps.org/doi/10.1103/PhysRevA.102.012206}
{Phys. Rev. A {\bf 102}, 012206 (2020)}.

\bibitem {Ming2020} 
F. Ming, X.-K. Song, J. Ling, L. Ye, and D. Wang, 
\href{https://doi.org/10.1140/epjc/s10052-020-7840-y}
{Eur. Phys. J. C {\bf 80}, 275 (2020)}.

\bibitem {PhysRevA.106.062219} 
 L. Wu, L. Ye, and D. Wang, \href{https://link.aps.org/doi/10.1103/PhysRevA.106.062219}
{Phys. Rev. A {\bf 106}, 062219 (2022)}.

\bibitem {PhysRevE.106.054107} 
M.-L. Song, L.-J. Li, X.-K. Song, L. Ye, and D. Wang,
\href{https://link.aps.org/doi/10.1103/PhysRevE.106.054107}
{Phys. Rev. E {\bf 106}, 054107 (2022)}.

\bibitem {Li2022} 
L.-J. Li, F. Ming, X.-K. Song, L. Ye, and D. Wang, 
\href{https://doi.org/10.1140/epjc/s10052-022-10687-1}
{Eur. Phys. J. C {\bf 82}, 726 (2022)}.

\bibitem {PhysRevA.104.062204} 
B.-F. Xie, F. Ming, D. Wang, L. Ye, and J.-L. Chen, 
\href{https://link.aps.org/doi/10.1103/PhysRevA.104.062204}
{Phys. Rev. A {\bf 104}, 062204 (2021)}.

\bibitem {Li2021} 
L.-J. Li, F. Ming, X.-K. Song, L. Ye, and D. Wang, 
\href{https://doi.org/10.1140/epjc/s10052-021-09503-z}
{Eur. Phys. J. C {\bf 81}, 728 (2021)}.

\bibitem {PhysRevA.101.032101} 
Z.-Y. Ding, H. Yang, D. Wang, H. Yuan, J. Yang, and L. Ye,
\href{https://link.aps.org/doi/10.1103/PhysRevA.101.032101}
{Phys. Rev. A {\bf 101}, 032101 (2020)}.

\bibitem {PhysRevE.109.064103} 
M.-L. Song, X.-K. Song, L. Ye, and D. Wang, \href{https://link.aps.org/doi/10.1103/PhysRevE.109.064103}
{Phys. Rev. E {\bf 109}, 064103 (2024)}.

\bibitem {PhysRevLett.113.140401} 
T. Baumgratz, M. Cramer, and M. B. Plenio, \href{https://link.aps.org/doi/10.1103/PhysRevLett.113.140401}
{Phys. Rev. Lett. {\bf 113}, 140401 (2014)}.

\bibitem {PhysRevLett.80.2245} 
W. K. Wootters,
\href{https://link.aps.org/doi/10.1103/PhysRevLett.80.2245}
{Phys. Rev. Lett. {\bf 80}, 2245 (1998)}.

\bibitem {Eisert_1999} 
J. Eisert and M. B. Plenio,
\href{http://dx.doi.org/10.1080/09500349908231260}
{J. Mod. Opt. {\bf 46}, 145-154 (1999)}.

\bibitem {PhysRevLett.103.160504} 
M. Piani,
\href{https://link.aps.org/doi/10.1103/PhysRevLett.103.160504}
{Phys. Rev. Lett. {\bf 103}, 160504 (2009)}.

\bibitem {PhysRevA.87.062103} 
J. Schneeloch, C. J. Broadbent, S. P. Walborn, E. G. Cavalcanti,
and J. C. Howell,
\href{https://link.aps.org/doi/10.1103/PhysRevA.87.062103}
{Phys. Rev. A {\bf 87}, 062103 (2013)}.

\bibitem {PhysicsPhysiqueFizika.1.195} 
J. S. Bell,
\href{https://link.aps.org/doi/10.1103/PhysicsPhysiqueFizika.1.195}
{Physics   {\bf 1}, 195 (1964)}.

\bibitem {PhysRevA.75.022108} 
S. Ashhab, K. Maruyama, and F. Nori,
\href{https://link.aps.org/doi/10.1103/PhysRevA.75.022108}
{Phys. Rev. A {\bf 75}, 022108 (2007)}.

\bibitem {PhysRevLett.23.880} 
J. F. Clauser, M. A. Horne, A. Shimony, and R. A. Holt,
\href{https://link.aps.org/doi/10.1103/PhysRevLett.23.880}
{Phys. Rev. Lett. {\bf 23}, 880 (1969)}.

\bibitem {PhysRevLett.88.017901} 
H. Ollivier and W. H. Zurek,,
\href{https://link.aps.org/doi/10.1103/PhysRevLett.88.017901}
{Phys. Rev. Lett. {\bf 88}, 017901 (2001)}.

\bibitem {PhysRevD.29.1047} 
W. G. Unruh and R. M. Wald, \href{https://link.aps.org/doi/10.1103/PhysRevD.29.1047}
{Phys. Rev. D {\bf 29}, 1047 (1984)}.

\bibitem {PhysRevD.68.085006} 
P. Kok and U. Yurtsever, \href{https://link.aps.org/doi/10.1103/PhysRevD.68.085006}
{Phys. Rev. D {\bf 68}, 085006 (2003)}.

\bibitem {PhysRevA.80.032315} 
A. G. S. Landulfo and G. E. A. Matsas,
\href{https://link.aps.org/doi/10.1103/PhysRevA.80.032315}
{Phys. Rev. A {\bf 80}, 032315 (2009)}.

\bibitem {PhysRevA.81.062130} 
L. C. C\'eleri, A. G. S. Landulfo, R. M. Serra, and G. E. A.
Matsas,
\href{https://link.aps.org/doi/10.1103/PhysRevA.81.062130}
{Phys. Rev. A {\bf 81}, 062130 (2010)}.

\bibitem {Tian_2012} 
Z. Tian, J. Wang, and J. Jing,
\href{http://dx.doi.org/10.1016/j.aop.2013.01.015}
{Ann. Phys. {\bf 332}, 98-109 (2012)}.

\bibitem {Wang2014} 
J. Wang, Z. Tian, J. Jing, and H. Fan,
\href{https://doi.org/10.1038/srep07195}
{Sci. Rep. {\bf 4}, 7195 (2014)}.

\bibitem {Wald:1995yp} 
R. M. Wald,
{\it Quantum Field Theory in Curved Space-Time
and Black Hole Thermodynamics, Chicago Lectures in Physics}
(University of Chicago Press, Chicago, IL, 1995).

\bibitem {Wang_2011} 
C.-Z. Wang, C.-X. Li, L.-Y. Nie, and J.-F. L,
\href{https://dx.doi.org/10.1088/0953-4075/44/1/015503}
{J. Phys. B: At., Mol. Opt. Phys. {\bf 44}, 015503 (2010)}.


\end{references}

\end{document}